\documentstyle[12pt]{article}
\begin{document}
\newcommand {\ber} {\begin{eqnarray*}}
\newcommand {\eer} {\end{eqnarray*}}
\newcommand {\bea} {\begin{eqnarray}}
\newcommand {\eea} {\end{eqnarray}}
\newcommand {\beq} {\begin{equation}}
\newcommand {\eeq} {\end{equation}}
\newcommand {\state} [1] {\mid \! \! {#1} \rangle}
\newcommand {\sym} {$SY\! M_2\ $}
\newcommand {\eqref} [1] {(\ref {#1})}
\newcommand{\preprint}[1]{\begin{table}[t] 
           \begin{flushright}               
           \begin{large}{#1}\end{large}     
           \end{flushright}                 
           \end{table}}                     
\def\Acknowledgements{\bigskip  \bigskip {\begin{center} \begin{large}
             \bf ACKNOWLEDGMENTS \end{large}\end{center}}}

\newcommand{\half} {{1\over {\sqrt2}}}
\newcommand{\dx} {\partial _1}

\def\Dslash{\not{\hbox{\kern-4pt $D$}}}
\def\cmp#1{{\it Comm. Math. Phys.} {\bf #1}}
\def\cqg#1{{\it Class. Quantum Grav.} {\bf #1}}
\def\pl#1{{\it Phys. Lett.} {\bf #1B}}
\def\prl#1{{\it Phys. Rev. Lett.} {\bf #1}}
\def\prd#1{{\it Phys. Rev.} {\bf D#1}}
\def\prr#1{{\it Phys. Rev.} {\bf #1}}
\def\pr#1{{\it Phys. Rept.} {\bf #1}}
\def\np#1{{\it Nucl. Phys.} {\bf B#1}}
\def\ncim#1{{\it Nuovo Cimento} {\bf #1}}
\def\lnc#1{{\it Lett. Nuovo Cim.} {\bf #1}}
\def\jmath#1{{\it J. Math. Phys.} {\bf #1}}
\def\mpl#1{{\it Mod. Phys. Lett.}{\bf A#1}}
\def\jmp#1{{\it J. Mod. Phys.}{\bf A#1}}
\def\aop#1{{\it Ann. Phys.} {\bf #1}}
\def\hpa#1{{\it Helv.Phys.Acta} {\bf#1}}
\def\mycomm#1{\hfill\break{\tt #1}\hfill\break}
\def\dslash{\not{\hbox{\kern-4pt $\partial$}}}
\begin{titlepage}
\rightline{TAUP-2566-99}
\rightline{WIS-99/09/Mar-DPP}
\rightline{\today}
\vskip 1cm
\centerline{{\Large \bf The String Tension in Two Dimensional \newline
    }}
\centerline{{\Large \bf Gauge Theories}}
\vskip 1cm
\centerline{A. Armoni$^a$
, Y. Frishman$^b$
 and J. Sonnenschein$^a$}
\begin{center}
\em $^a$School of Physics and Astronomy
\\Beverly and Raymond Sackler Faculty of Exact Sciences
\\Tel Aviv University, Ramat Aviv, 69978, Israel
\\and
\\ $^b$Department of Particle Physics
\\Weizmann Institute of Science
\\76100 Rehovot, Israel
\end{center}

\begin{abstract}
We review and elaborate on  properties of the 
string tension in two-dimensional gauge theories.

The first model we consider is  massive QED  in the  $m\ll e$ limit. 
We evaluate the leading string tension both in
the fermionic and bosonized descriptions. 
We discuss the next to leading corrections in $m/e$. 
The next-to-leading terms in the long distance behavior of the
quark-antiquark
potential, are evaluated in a certain region of external versus
dynamical charges. The finite temperature behavior is also determined.  

In $QCD_2$ we review the results for  the  string tension of  quarks
in cases with dynamical quarks in the fundamental, adjoint, symmetric
and antisymmetric representations. 
The screening nature of $SYM_2$ is re-derived. 

\end{abstract}
\end{titlepage}
\section{Introduction}
Two-dimensional gauge theories serve as a theoretical laboratory
for studying four-dimensional gauge theories. Non-perturbative issues, such as
confinement and spectrum of models, can be addressed in these theories.
For recent reviews see\cite{FSrev,Abdalla}.

In this framework, it is also possible to calculate the string tension
$\sigma$ of
the confining part of the potential,
\beq
V=\sigma r \label{potential}
\eeq  
 
The leading term of the string tension (in mass over charge
parameter), in the massive Schwinger model 
($U(1)$ gauge theory with massive matter, which we call electron), 
was calculated by using bosonization, long time ago\cite{CJS}
\beq
\sigma _{QED} = m\mu \left (1-\cos (2\pi {q_{ext} \over q_{dyn}})
\right )  \label{abelian},
\eeq
where $m$ is the electron mass, $\mu = e{\exp (\gamma) \over
  2\pi^{3/2}}$, $e$ the gauge coupling, $\gamma$ the Euler number
 and $q_{ext}$,$q_{dyn}$ are the external and dynamical charges
respectively (we measure charges in units of $e$, thus $q_{ext}$ and
$q_{dyn}$ are dimensionless).

Note that the string tension vanishes whenever the external charge is an integer
multiple of the dynamical charge, $q_{ext}= n q_{dyn}$. This is
expected, since in this case dynamical charges, via electron-positron
pairs, can screen the external source. Explicitly, $n$ pairs are
created, with $n$ electrons screening the positive charge and the $n$
positrons the negative one. Another important and somewhat unexpected
result is that the string tension vanishes also when $m=0$. This
phenomenon can be explained in several ways. In the massless theory it
is easy to produce pairs from the vacuum. Therefore, infinite amount
 of integer charges which are produced, may form a coherent state with a
fractional charge and screen the fractional external charge. A second
explanation is that, due to the chiral anomaly, the photon becomes
massive resulting in a short range potential.

The expression in massive $QCD_2$ is\cite{AFS}
\beq
\sigma _{QCD} =  m \mu _R \
\sum _i \left ( 1-\cos (4\pi \lambda _i  {k_{ext}  \over k_{dyn}}
  )\right )  \label{nonabelian}
\eeq
where  $\mu _R = e {\exp (\gamma) \over (2\pi)^{3/2}}$,
 $\lambda _i$ are the isospin eigenvalues of the
dynamical representation, $k_{ext}$ and $k_{dyn}$ are the affine
current algebra levels of the external and dynamical representations, respectively. This
expression was shown to hold for the fundamental and the adjoint
representations. Other representations were also discussed
in\cite{AFS}, with appropriate expressions for the string tension, as
generalizations of eq.\eqref{nonabelian}.

Note that when $m=0$ the string tension
\eqref{nonabelian} vanishes, as in the Abelian case\cite{gross}. The explanation via
acquiring mass is now more involved than in QED, as  we are now in the gauge dependent
sector\cite{FS,dalley}.
Another explanation, which has no direct Abelian analogue, is related to the
equivalence theorem of Kutasov and Schwimmer\cite{KS}. The massless
adjoint fermions model is physically equivalent to the multi-flavor
massless model with $N_f=N_c$ fermions in the fundamental
representation.
Therefore the original adjoint fermion can be expressed as a 
fundamental  fermions which can screen the external source.

The plan of the paper, which is an expanded version of\cite{AFS} and
\cite{AFS2}, but also with several new results, is as follows. In section 2 and 3 we calculate the string
tension for the massive Schwinger model in both the fermionic
and the bosonic languages. The bosonic language will be useful for
the non-Abelian generalization and fermionic language will be useful
when we will discuss supersymmetric theories.  

Section 4 is devoted to quantum corrections to the string
tension. Note that the expressions \eqref{abelian} and
\eqref{nonabelian} are only the leading terms in mass perturbation
theory and are valid when $m\ll e$. The next to leading order
correction, in the Abelian case, was derived in\cite{Adam} and it is
reviewed briefly in this section.

In section 5 we discuss the short range corrections to the
confining potential. We focus on the Abelian case, believing
that the non-Abelian case is very similar. Our conclusion is that apart
from the linear potential,  a screening part, which arise from a massive
component of the photon/gluon, is present.

In section 6 we comment on the behavior of the string tension when
finite temperature is introduced. We follow ref.\cite{RH} and conclude
that confinement persists even at high temperatures. This is peculiar to two dimensions.

Section 7 and 8 are devoted to the non-Abelian generalization. We
compute the string tension for the cases of matter in the fundamental
and adjoint representations (section 7) and symmetric and
anti-symmetric representations (section 8). These sections are based
on ref.\cite{AFS}.

In section 9 we show that the string tension vanishes in
supersymmetric gauge theories by showing that there is no 
$<tr \phi \bar \lambda \gamma _5 \lambda >= 0$
condensate in these models.

The appendix is devoted to a derivation of the quark anti-quark
external current. It is shown that the relevant charge of the external
source is the chiral anomaly (the affine Lie algebra level).

\section{The Schwinger model}

Let us review the derivation of the string tension in the massive
Schwinger model, in the fermionic language. Consider the partition function of two dimensional massive QED
\bea 
\lefteqn{Z=} \label{partition} \\ &&
\int DA_\mu D\bar \Psi D \Psi \exp \left ( i\int d^2x\ (-{1\over 4e^2} F_{\mu\nu}^2
+\bar \Psi i\dslash \Psi - m\bar \Psi \Psi - q_{dyn}A_\mu \bar \Psi \gamma ^\mu
\Psi )\right ), \nonumber
\eea
where $q_{dyn}$ is the charge of the dynamical fermions. Gauge fixing
terms were not written explicitly.
Let us add an external pair with charges $\pm q_{ext}$
at $\pm L$, namely $j_0^{ext} = q_{ext}(\delta (x+L) - \delta
(x-L))$, so that the change of ${\cal L}$ is $-j_\mu ^{ext} A^{\mu}
(x)$. Note that by choosing $j_\mu ^{ext}$ which is conserved,
$\partial ^\mu j^{ext}_\mu =0$, the action including the coupling to the
external current is also gauge invariant.

Now, one can eliminate this charge by performing a local, space-dependent
 left-handed rotation
\bea
 \Psi \rightarrow e^{i\alpha(x) {1\over 2}(1-\gamma_5) } \Psi \\
  \bar \Psi \rightarrow \bar \Psi e^{-i\alpha(x) {1\over 2}(1+\gamma_5) } , 
\eea
where $\gamma ^ 5 = \gamma ^0 \gamma ^1$.
 We choose a left-handed rotation (or equally well a right-handed one)
 rather than an axial one, in analogy with the non-abelian case (see
 section 7), where it is simpler to do so.

The rotation introduce a change in the action, due to the chiral anomaly
\bea
\delta S = \int d^2 x {\alpha(x) q_{dyn} \over 2\pi} F,
\eea
where $F$ is the dual of the electric field $F={1\over 2} \epsilon
^{\mu \nu} F_{\mu \nu}$.

The new action is 
\bea \label{rotated}
\lefteqn{S=} \\ &&
\int d^2x\ (-{1\over 4e^2} F_{\mu\nu}^2
+\bar \Psi i\dslash \Psi -\bar \Psi \partial_\mu \alpha(x) \gamma ^\mu
{1\over 2}(1-\gamma _5) \Psi 
 - m\bar \Psi e^{-i\alpha(x) \gamma_5 }\Psi  \nonumber \\ &&
 - q_{dyn}A_\mu \bar \Psi \gamma ^\mu\Psi - q_{ext}(\delta (x+L) -
 \delta (x-L))A_0 +  {\alpha(x) q_{dyn} \over 2\pi} F) \nonumber 
\eea

The external source and the anomaly term are similar, both being
linear in the gauge potential. This is the
reason that the $\theta$-vacuum and electron-positron pair at the boundaries are the same in
two-dimensions\cite{CJS}. In the following we assume $\theta =0$, as
otherwise we absorb it in $\alpha $. Choosing the $A_1=0$ gauge and integrating by
parts, the anomaly term looks like an external source
\beq
{q_{dyn}\over 2\pi} A_0 \partial_1 \alpha(x)
\eeq 
This term can cancel the external source by the choice
\beq
  \alpha(x) = 2\pi {q_{ext} \over q_{dyn}} (\theta (x+L) -\theta
   (x-L)).
\eeq 
Let us take the limit $L\rightarrow \infty$.
The form of the action, in the region $B$ of $-L<x<L$ is
\beq
S_B =\int _B d^2x\ (-{1\over 4e^2} F_{\mu\nu}^2
+\bar \Psi i\dslash \Psi 
 - m\bar \Psi e^{-i 2\pi {q_{ext} \over q_{dyn}} \gamma_5 }\Psi
 - q_{dyn}A_\mu \bar \Psi \gamma ^\mu\Psi ) \label{rotated2}
\eeq
Thus the total impact of the external electron-positron pair is a chiral
rotation of the mass term. This term can be written as
\beq
\bar \Psi e^{-i 2\pi {q_{ext} \over q_{dyn}} \gamma_5 }\Psi=
\cos  (2\pi {q_{ext} \over q_{dyn}}) \bar \Psi \Psi -  i\sin (2\pi {q_{ext} \over q_{dyn}})
\bar \Psi \gamma_5 \Psi
\eeq
The string tension is the vacuum expectation value (v.e.v.) of the
Hamiltonian density in the presence of the external source relative to the
v.e.v. of the Hamiltonian density without the external source, in the
$L\rightarrow \infty$ limit.  
\beq
\sigma = <{\cal H}>-<{\cal H}_0>_0 \label{diff}
\eeq
 where $\state{0}_0 $ is the vacuum state with no external sources.
The change in the vacuum energy is due to the mass term. The change in
the kinetic term which appears in \eqref{rotated} does not contribute to the
vacuum energy\cite{AFS}.
Thus
\beq
\sigma = 
m\cos  (2\pi {q_{ext} \over q_{dyn}}) <\bar \Psi \Psi> -m \sin (2\pi {q_{ext} \over q_{dyn}})
<\bar \Psi i\gamma_5 \Psi> -m<\bar \Psi \Psi>_0
\eeq
Thus, the values of the condensates $<\bar \Psi \Psi>$  and $<\bar
\Psi \gamma_5 \Psi>$ are needed. The easiest way to compute these
condensates is Bosonization, but it can also be computed directly in
the fermionic language which at $m=0$ is \cite{JSW}
\bea
 && <\bar \Psi \Psi>_{m=0} =  -e{\exp (\gamma) \over 2\pi^{3/2}}
 \label{condensate} \\
&& <\bar \Psi \gamma_5 \Psi>_{m=0} =0 ,\label{trivial}
\eea

Eq.\eqref{trivial} is due to parity
invariance (with our choice $\theta =0$). The resulting string tension, to first order in $m$,
\beq 
\sigma = m  e{\exp (\gamma) \over 2\pi^{3/2}} \left (1-\cos  (2\pi {q_{ext}
  \over q_{dyn}})\right ) \label{fermtension}
\eeq

Though this expression is only the leading term in a $m/e$
expansion and might be corrected\cite{Adam}, when $q_{ext}$ is an integer multiple of
$q_{dyn}$ the string tension is {\em exactly} zero, since in this case
the rotated action\eqref{rotated2} is not changed from the original
one \eqref{partition}.

\section{The Schwinger model in Bosonic form}
In their seminal paper, Coleman Jackiw and Susskind used\cite{CJS} the bosonized
version of the Schwinger model to calculate the string tension. We
present here their calculation, for completeness.

The bosonized Lagrangian, in the gauge $A_1=0$, is the following
\beq
{\cal L} = {1\over 2e^2} (\partial _1 A_0)^2 + {1\over 2} (\partial _\mu \phi)^2
+ M^2 \cos (2\sqrt \pi\phi) + {q_{dyn}\over \sqrt \pi} A_0 \partial _1 \phi - A_0 j_{ext},
\label{bosschwinger}
\eeq
where $M^2 =m \mu$, $\mu={{\exp (\gamma)}\over 2\pi} \mu_{(\phi)}$,
with $\mu _{(\phi)} = {e\over \sqrt \pi} q_{dyn}$ the mass of the
photon for $e \gg m$.

 Chiral rotation corresponds to a shift in the field $\phi$. Upon the
 transformation
\beq
\phi =  \tilde \phi + \sqrt \pi {q_{ext} \over q_{dyn}} \left (\theta
  (x+L)- \theta (x-L) \right ),
\eeq
The Lagrangian \eqref{bosschwinger} takes, in the region $B$,  the form
 \beq
{\cal L} _B = {1\over 2e^2} (\partial _1 A_0)^2 + {1\over 2} (\partial
_\mu \tilde \phi)^2
+ M^2 \cos (2\sqrt \pi\tilde \phi+2\pi {q_{ext}\over q_{dyn}}) +
{q_{dyn}\over \sqrt \pi} A_0 \partial _1
\tilde \phi 
\label{bosschwinger2}
\eeq
Hence, similarly to the previous derivation, a local chiral rotation
may be used to eliminate the external source. The calculation of the
string tension is exactly the same as in the previous section.

The relevant part of the Hamiltonian density is 
\beq 
{\cal H} = - M^2 \cos (2\sqrt \pi\tilde \phi+2\pi {q_{ext}\over
  q_{dyn}}) \label{bosHam}
\eeq   
 To zeroth order in ${({M \over e})}^2$, the vacuum is
$\tilde \phi =0$. Setting this choice in \eqref{bosHam} and subtracting the v.e.v. of the free
Hamiltonian, we arrive at \eqref{abelian}.

\section{Beyond the small mass Abelian string tension}

The expression \eqref{abelian} contains only the leading $m\over
e$ contribution to the Abelian string tension. This expression was
computed in section 3, using a classical average.
However, as we used the normal ordering scale $\mu _\phi$ which is
the photon mass for $e\gg m$, taking $\tilde \phi =0$ actually gives
the full quantum answer, as is evident by comparing with the fermionic
calculation of section 2.

The full perturbative (in $m$) string tension can be written as \cite{Adam}
\beq
\sigma _{QED} = m \mu \sum _{l=1} ^{\infty}  C_l 
( {m\over e q_{dyn}})^{l-1} \left(1-\cos
(2\pi l {q_{ext}\over q_{dyn}})\right ) \label{fullabelian}    
\eeq

The value of the first coefficient is $C_1=1$ and the next is $C_2
=-8.91{\exp (\gamma) \over 8\pi ^{1/2}}$ \cite{Adam2}. Higher
coefficients were not calculated yet.

Note that for finite ${m\over e}$ we have to minimize the potential
\beq
 V= M^2 \left (1-\cos (2\sqrt \pi \phi + 2\pi {q_{ext}\over q_{dyn}}) \right ) + {1\over 2} \mu _
 {\phi} ^2 \phi ^2 .
\eeq
The minimum $\phi = \phi _m$ obeys
\beq
 2\sqrt \pi M^2 \sin (2\sqrt \pi \phi_m + 2\pi {q_{ext}\over q_{dyn}}) + \mu _ \phi ^2 \phi _ m =0
\eeq
Thus, for the first order $({m\over e q_{dyn}})$ correction, we get a 
$C_2$ which is  
$-({1\over 2}) {\sqrt \pi} ({\exp \gamma})$. This has the 
same sign, but a factor 1.41 larger, than the instanton contribution ref (13).

Note that all above results for the string tension are symmetric under
change of sign of the external charge, as expected on general
grounds. However, when a $\theta F$ term is introduced,
 we get odd terms as well, like
 $\sin (l\theta) \sin (2\pi l{q_{ext} \over q_{dyn}})$ \cite{Adam}. The even
terms are multiplied by $\cos (l\theta)$.

We expect that similar corrections as those in eq.\eqref{fullabelian} will occur in the non-Abelian
case. For the fundamental/adjoint case, the following expression
may correct the leading term\eqref{nonabelian}
\beq
\sigma _{QCD} =  m \mu _R \sum _{l=1} ^{\infty} {\tilde C}_l ({m\over
  e k_{dyn}})^{l-1}
\sum _j \left ( 1-\cos (4\pi \lambda _j l {k_{ext}  \over k_{dyn}}
  )\right )  
\eeq

Finally, let us remark that for very large ${m\over e}$, the abelian
case has a string tension which is ${1\over 2} e^2 q_{ext}^2$.

\section{Correction to the leading long distance Abelian potential}
The potential \eqref{potential} is the dominant long range
term. However, there are, of course, corrections. In this section we
present these corrections.

The equations of motions which follow from the bosonized Lagrangian
\eqref{bosschwinger} are, in the static case
\bea
&& -{1\over e^2} \partial _1^2 A_0 + {q_{dyn} \over \sqrt \pi} \partial _1 \phi -j_{ext}
=0 \label{bos1} \\
&& -\partial _1 ^2 \phi + 2\sqrt \pi M^2 \sin 2\sqrt \pi \phi +
{q_{dyn} \over \sqrt \pi} \partial _1 A_0 =0 \label{bos2}
\eea
In order to solve these equation, it is useful to eliminate the
bosonized matter field $\phi$. Using the approximation $\sin 2\sqrt
\pi \phi \sim 2\sqrt \pi \phi$, we arrive at (in momentum space),
\beq 
A_0 (k) = 
{{e^2(k^2+4\pi M^2)}\over {k^2(k^2+(4\pi M^2+{e^2\over \pi}
    q_{dyn}^2))}}j_{ext}(k)
\eeq
where $k$ is the Fourier transform of the space coordinate. We will
discuss the validity of our approximation for $\phi$ later in this
section. The last equation can be rewritten as 
\beq
A_0(k) =
 \left ( {m_1^2 \over m_2 ^2} {1\over k^2} + (1-{m_1^2\over
    m_2^2}){1\over {k^2 + m_2 ^2}} \right )e^2 j_{ext}(k) \nonumber
\eeq
where 
\bea
&& m_1 ^2 = 4\pi M^2 \\
&& m_2^2 = 4\pi M^2 + {e^2 \over \pi} q_{dyn}^2
\eea

Note that the photon propagator has two poles; a massless pole
reproduces the string tension and a  massive pole which
 adds a screening term to the potential. Note that there is no
 ${const.\over L}$ correction, which appears in higher dimensions\cite{LSW}, since
 in the present case the string cannot fluctuate in transverse directions.

Note also that in the massless case, when $M^2=0$, only the second
term survives and the photon has only one pole with mass square ${e^2
  \over \pi} q_{dyn}^2$. This result is of course exact, independent
of our approximation. 

The resulting gauge field is
\bea
\lefteqn { A_0(x)= } \\ && 
{2\pi^2 M^2 q_{ext} \over  q_{dyn}^2} \left ( \mid x+L \mid
  - \mid x-L \mid \right ) \nonumber \\ && 
- {e\sqrt \pi \over 2} {q_{ext} \over q_{dyn}}\left ( e^{-{e\over \sqrt \pi} q_{dyn}
    \mid x+L \mid} - e^{-{e\over \sqrt \pi} q_{dyn}
    \mid x-L \mid} \right ) \nonumber
\eea
where we took $M^2 \ll e^2$ for simplicity. 

In order to calculate the potential we will use
\beq 
V= {1\over 2} \int A_0(x) j_{ext}(x) dx
\eeq
Hence the potential is
\beq
V= 2 \pi ^2 M^2 {q_{ext}^2 \over q_{dyn}^2} \times 2L
   + {e \sqrt \pi \over 2} {q_{ext}^2 \over q_{dyn}}(1-e^{-{e\over \sqrt \pi} q_{dyn} 2L})
\eeq
The first term is the confining potential which exists whenever the
quark mass is non-zero. On top of it, there is always a screening potential.

The string tension which results from the above potential is
\beq
\sigma =  m\mu \times 2 \pi ^2 {q_{ext}^2 \over q_{dyn}^2}
\eeq 
which is exactly \eqref{abelian} in the approximation $2\pi
{q_{ext}\over q_{dyn}} \ll 1$. This turns out to be also the condition
for $\sin 2\sqrt \pi \phi \sim 2\sqrt\pi \phi$ that we assumed in the
start of this section. To see that, we solve for $\phi$ from
eq.\eqref{bos1} as 
\beq
\phi(k) = -ik {q_{dyn} \over \sqrt \pi} {e^2 \over m_2 ^2} ({1\over
  k^2} - {1\over k^2 + m_2 ^2} ) j_{ext}(k)
\eeq
Define $\phi = \phi_1 + \phi_2$, where $\phi_1$ is the part with
${1\over k^2}$, and $\phi_2$ with ${1\over k^2 + m_2 ^2}$. The
$\phi_2$ part goes to zero at long distances, i.e. $k\rightarrow 0$. As
for the $\phi_1$ part, its x-space form is 
\beq 
 \phi_1(x) = {e^2 \over \sqrt \pi m_2 ^2} q_{dyn} q_{ext} (\theta
 (x+L) - \theta (x-L))
\eeq
which for small ${m\over e}$ reduced to
\beq
\phi_1 (x) \sim \sqrt \pi {q_{ext} \over q_{dyn}} (\theta (x+L)  -
\theta (x-L))
\eeq
Thus $2\sqrt \pi \phi$ small means
\beq
(2\pi) {q_{ext} \over q_{dyn}} \ll 1
\eeq
the condition mentioned before.

Note that we could generalize the argument to values of $ 2\pi
{q_{ext} \over q_{dyn}}$ that are close to $2\pi n$, with integer $n$. 
\section{Finite temperature}
In this section we would like to comment on the behavior of the string
tension in the presence of finite temperature. It is interesting to
check whether the string is torn due to high temperature and whether
the system undergoes a phase transition from confinement to
de-confinement.

The prescription for calculating quantities at finite temperature $T$
 is to formulate the theory on a circle in Euclidean time with circumference
$\beta = T^{-1}$.

For the purpose of calculating the string tension, we can follow the same
steps which were used in sections 2 and 3 leading to a modification of
eq.\eqref{fermtension} as (a comprehensive discussion of this issue is given at\cite{RH})
\beq 
\sigma = -m <\bar \Psi \Psi> _T (1-\cos 2\pi {q_{ext} \over q_{dyn}})
\eeq
It is enough to calculate $<\bar \Psi \Psi>_T$, the condensate at finite
temperature, in the massless Schwinger model. 

Following ref.\cite{JSW}, the chiral condensate behaves as 
\beq
<\bar \Psi \Psi>_{(T\rightarrow 0)} \rightarrow -{e \over 2\pi^{3/2}} e^\gamma ,
\eeq 
 and 
\beq
<\bar \Psi \Psi>_{ (T\rightarrow \infty)} \rightarrow -2T e^{-{\pi
    ^{3/2} T \over e}} \label{highT} .
\eeq 

The above result \eqref{highT} indicates that the string is not torn
even at very high temperatures. The explicit expression in\cite{RH}
shows that $<\bar \Psi \Psi>_T$ is non-zero for all $T$. Thus, the system does not undergoes a
phase transition.  It is just energetically favorable to have the
electron-positron pair confined.

\section{Two-dimensional QCD}
The action of bosonized $QCD_2$ with massive quarks in the fundamental
representation of $SU(N)$ is \cite{FSrev}
\bea
\lefteqn{S_{fundamental}={1\over{8\pi}}\int _\Sigma d^2x \ tr(\partial _\mu
g\partial ^\mu g^\dagger) + } \label{fund} \\
 && {1\over{12\pi}}\int _B d^3y \epsilon ^{ijk} \
 tr(g^\dagger\partial _i g) (g^\dagger\partial _j g)(g^\dagger\partial _k g) +
 \nonumber \\
&& {1\over 2} m \mu _{fund} \int d^2x \ tr (g+g^\dagger)  
-\int d^2 x{1\over {4e^2}} F^a _{\mu \nu} F^{a \mu \nu} - \nonumber \\
&& {1\over 2\pi}\int d^2 x \ tr (ig^\dagger \partial_+ g A_-
+ig\partial_ - g^\dagger A_+ + A_+ g A_- g^\dagger - A_+ A_-), \nonumber
\eea
where $e$ is the gauge coupling, $m$ is the quark mass, $\mu = e {\exp
  (\gamma ) \over (2\pi)^{3\over 2}}$,
 $g$ is an $N\times N$ unitary matrix, $A_\mu$ is the gauge field and the
trace is over $U(N)$ indices. Note, however, that only
 the $SU(N)$ part of the matter field $g$ is gauged.

When the quarks transform in the adjoint representation, the
expression for the action is\cite{AGSY}
\bea
\lefteqn{S_{adjoint}={1\over{16\pi}}\int _\Sigma d^2x \ tr(\partial _\mu
g\partial ^\mu g^\dagger) + } \label{adjoint} \\
 && {1\over{24\pi}}\int _B d^3y \epsilon ^{ijk} \
 tr(g^\dagger\partial _i g) (g^\dagger\partial _j g)(g^\dagger\partial
 _k g) +
 \nonumber \\
&& {1\over 2} m \mu _{adj} \int d^2x \ tr (g+g^\dagger)  
-\int d^2 x{1\over {4e^2}} F^a _{\mu \nu} F^{a \mu \nu} - \nonumber \\
&& {1\over 4\pi}\int d^2 x \ tr (ig^\dagger \partial_+ g A_-
+ig\partial_ - g^\dagger A_+ + A_+ g A_- g^\dagger - A_+ A_-) \nonumber
\eea 
A version which takes into account
instanton effects is given in \cite{smilga}, but for our purposes it
will not be needed.

The action \eqref{adjoint} differs from \eqref{fund} by a factor of one half in front of the
 WZW and interaction terms, because $g$ is real and represents Majorana
fermions. Another difference is that $g$ now is an
$(N^2-1)\times(N^2-1)$ orthogonal matrix. The two actions 
\eqref{fund} and \eqref{adjoint} can be schematically
 represented by one action 
\bea
 &&  S=S_0 + {1\over 2} m \mu _R \int d^2x \ tr (g+g^\dagger)
 + \label{universal} \\
&&
 -{i k_{dyn}\over 4\pi}\int d^2 x \ (g\partial_ - g^\dagger)^a A_+^a, \nonumber
\eea
 where $A_- = 0$ gauge was used, $S_0$ stands for the WZW action and
 the kinetic
 action of the gauge field, $k_{dyn}$ is the level (the chiral
 anomaly) of the dynamical
 charges ($k=1$ for the fundamental representation of $SU(N)$ and $k=N$
 for the adjoint representation).

Let us add an external charge to the action. We choose static
(with respect to the light-cone coordinate $x^+$) charge and
therefore we can omit its kinetic term
from the action. Thus an external charge coupled to the gauge field
would be represented by
\[
-{i k_{ext}\over 4\pi}\int d^2 x \ (u\partial_ - u^\dagger)^a A_+^a 
\]
Suppose that we want to put a quark and an anti-quark at a very large
separation. A convenient choice of the charges would be a
direction in the algebra in which the generator has a diagonal
form. The simplest choice is a generator of an $SU(2)$
subalgebra. Since a rotation in the algebra is always possible, the results are
insensitive to this specific choice. As an
example we write down the generator in the case of fundamental and
adjoint representations.
\ber
 && T^3_{fund} = diag ({1\over 2},-{1\over 2},\underbrace{0,0,...,0}_{N-2}) \\
 && T^3_{adj} = diag (1,0,-1,
\underbrace{{1\over 2},-{1\over 2},{1\over 2},-{1\over 2},...,{1\over
    2},-{1\over 2}}_{2(N-2)\,\,\, doublets},\underbrace{0,0,...,0}_{(N-2)^2})
\eer  
Generally $T^3$ can be written as
\[
 T^3 = diag (\lambda _1,\lambda _2,...,\lambda
 _i,...,0,0,...), 
\]
where $\{ \lambda _i \}$ are the 'isospin'
 components of the representation under the
$SU(2)$ subgroup.

We take the $SU(N)$ part of $u$ as (see Appendix)
\beq
\label{external-charge}
u=\exp -i4\pi \left ( \theta (x^-+L )-\theta
  (x^--L )\right ) T^3_{ext},
\eeq
for $N>2$ and similar expression with a $2\pi$ factor 
for $N=2$. $T^3_{ext}$ represents the '3' generator of the external charge and
$u$ is static with respect to the light-cone time coordinate $x^+$.
The theta function is used as a limit of a smooth
function which interpolates between 0 and 1 in a very short
distance. In that limit $u=1$ everywhere except at isolated points,
where it is not well defined.  

The form of the action \eqref{universal} in the presence of an external
source is 
\ber
\lefteqn{
S=S_0 + {1\over 2} m \mu _R \int d^2x \left \{ \ tr
  (g+g^\dagger) +\right . } \\
&&
  \left . [ -{i k_{dyn}\over 4\pi} (g\partial_ - g^\dagger)^a
   +  k_{ext} \delta^{a3} ( \delta (x^-+L)-\delta
   (x^--L))] A_+^a \right \} 
\eer
The external charge can be eliminated from the action by a
transformation of the matter field. A new field $\tilde g$ can be
defined as follows
\ber
\lefteqn{
-{i k_{dyn}\over 4\pi} (\tilde g\partial_ - \tilde
g^\dagger)^a=} \\
&&
-{i k_{dyn}\over 4\pi} (g\partial_ - g^\dagger)^a 
+k_{ext} \delta^{a3}\left ( \delta (x^-+L)-\delta
   (x^--L) \right )   
\eer
This definition leads to the following equation for $\tilde g^\dagger$
\bea 
\lefteqn{\partial _- \tilde g^\dagger = }  \label{tilde} \\
&& \tilde g^\dagger \left (g\partial _-
  g^\dagger
+i4\pi {k_{ext} \over k_{dyn}} (\delta (x^-+L)-\delta
   (x^--L) ) T^3_{dyn} \right ) \nonumber
\eea
 The solution of \eqref{tilde} is 
\ber 
\lefteqn{
 \tilde g^\dagger =P \exp \left \{ \right . } \\
&&   \left . \int dx^- \left ( g\partial _-  g^\dagger
 +i4\pi {k_{ext} \over k_{dyn}} (\delta (x^-+L)-\delta
   (x^--L) ) T^3_{dyn} \right ) \right \}  \\
 &&  =e^{i4\pi {k_{ext} \over k_{dyn}}\theta(x^-+L) T^3_{dyn}}
 g^\dagger
e^{-i4\pi {k_{ext} \over k_{dyn}}\theta(x^--L) T^3_{dyn}} , 
\eer
where $P$ denotes path ordering and we assumed that $T^3_{dyn}$
commutes with $g\partial _- g^\dagger$ for $x^- \ge L$ and with $g^\dagger$
for $x^-=-L$ (as we shall see, this assumption is self consistent
with the vacuum configuration).

 Let us take the limit $L\rightarrow \infty$. For $-L<x^-<L$, the above relation simply means that
\[
g=\tilde g  e^{i4\pi {k_{ext} \over k_{dyn}} T^3_{dyn}}  
\]

Since the Haar
measure is invariant (and finite, unlike the fermionic case) with respect to unitary transformations,
the form of the action in terms of the new variable $\tilde g$ reads
\bea
\lefteqn{
S= 
 S_{WZW}(\tilde g) + S_{kinetic}(A_{\mu}) 
-{i k_{dyn}\over 4\pi}\int d^2 x \ (\tilde g\partial_ - \tilde
g^\dagger)^a A_+^a } \label{rotate} \\
&&
+ {1\over 2} m \mu _R \int d^2x \
 tr (\tilde g  e^{i4\pi {k_{ext}  \over k_{dyn}} T^3_{dyn}}   
+ e^{-i4\pi {k_{ext}  \over k_{dyn}} T^3_{dyn}} \tilde g^\dagger ) 
  \nonumber
\eea
which is $QCD_2$ with a chiraly rotated mass term.

The string tension can be calculated easily from \eqref{rotate} \cite{CJS}. It is
simply the vacuum expectation value (v.e.v.) of the Hamiltonian density,
relative to
the v.e.v. of the Hamiltonian density of the theory without an external source,
\[
\sigma = <H>-<H_0> 
\]
The vacuum of the theory is given by $\tilde g =1$. In terms of the
variable $g$, this configuration points in the '3' direction and hence
satisfies our assumptions while solving eq.\eqref{tilde}. The v.e.v. is
\ber
\lefteqn{
<H>= } \\
&&
-  {1\over 2} m \mu _R \
 tr ( e^{i4\pi {k_{ext}  \over k_{dyn}} T^3_{dyn}}   
+ e^{-i4\pi {k_{ext}  \over k_{dyn}} T^3_{dyn}} )
=  \\
&&
 -   m \mu _R \
\sum _i \cos (4\pi \lambda _i  {k_{ext}  \over k_{dyn}} ) 
\eer 
Therefore the string tension is
\beq
\sigma =  m \mu _R \
\sum _i \left ( 1-\cos (4\pi \lambda _i  {k_{ext}  \over k_{dyn}}
  )\right ) \label{sigma}
\eeq

which is the desired result.

A few remarks should be made:

(i) The string tension \eqref{sigma} reduces to the abelian string tension
\eqref{abelian} when abelian charges are considered. It follows that the
non-abelian generalization is realized by replacing the charge $q$
with the level $k$.

(ii) The string tension was calculated in the tree level of the
bosonized action. Perturbation theory (with $m$ as the
coupling) may change eq.\eqref{sigma}, since the loop effects may
add $O(m^2)$ contributions. However, we believe that it would
not change its general character. In fact, one feature is that the string tension vanishes
for any $m$ when ${k_{ext} \over k_{dyn}}$ is an integer, as follows
from eq.\eqref{rotate}, since the action does not depend then on
$k_{ext}$ at all.
 
(iii) When no dynamical mass is present, the theory exhibits screening. This
is simply because non-abelian charges at the end of the world interval
 can be eliminated from the action by
a chiral transformation of the matter field.

(iv) When the test charges are in the adjoint representation
$k_{ext} = N$,  equation \eqref{sigma} predicts screening by the fundamental
charges (with $k_{dyn} =1$).

(v) String tension appears when the test charges are in the fundamental
representation and the dynamical charges are in the adjoint
\cite{witten}. The value of the string tension is
\beq
\sigma = m \mu _{adj} \left (
2(1-\cos {4\pi \over N}) + 4(N-2)(1-\cos {2\pi \over N}) \right)
\eeq
 as follows from eq.\eqref{rotate} for this case.

 The case of $SU(2)$ is special. The $4\pi$ which
  appears in eq.\eqref{sigma} is replaced by $2\pi$, since the
 bosonized form of the external $SU(2)$ fundamental matter differs by
 a factor of a half with respect to the other $SU(N)$ cases
 (see Appendix). Hence, the string tension in this case is $4m \mu _{adj}$.

(vi) We would like to add, that when computing the string tension in
the pure YM case with external sources in representation $R$, the
Wilson loop gives ${1\over 2} e^2 C_2 (R)$, while our way of defining
external source gives ${1\over 2} e^2 k_{ext} ^2$. Thus we need a
factor ${C_2 (R) \over k _{ext} ^2}$ to bring our result to the Wilson
loop case. Analogous factors should be computed for the other cases,
when dynamical matter is also present.

\section{Symmetric and anti-symmetric representations}

The generalization of \eqref{sigma} to arbitrary representations is
not straightforward. However, we can comment about its nature (without
rigorous proof).

Let us focus on the interesting case of the antisymmetric
representation. One can show that in a similar manner to \cite{AGSY},
the WZW action with $g$ taken to be ${1\over 2}N(N-1) \times {1\over
  2}N(N-1)$ unitary matrices, is a bosonized version of $QCD_2$ with
fermions in the antisymmetric representation. 

The antisymmetric representation is described in the
Young-tableaux notation
 by two vertical boxes. Its dimension is ${1\over 2}N(N-1)$ and its diagonal
$SU(2)$ generator is
\bea
T^3_{as} = diag (
\underbrace{{1\over 2},-{1\over 2},{1\over 2},-{1\over 2},...,{1\over
    2},-{1\over 2}}_{(N-2)\,\,\, doublets},0,0,...,0),
\eea
 and consequently $k= N-2$. When the dynamical charges are
 in the fundamental and the external in the antisymmetric the string
 tension should vanish because tensor product of two fundamentals
 include the antisymmetric representation. Indeed,
 \eqref{sigma} predicts this result.

The more interesting case is when the dynamical charges are
antisymmetric and the external are fundamentals. In this case the
value of the string tension depends on whether $N$ is odd or even\cite{witten}.
When $N$ is odd the string tension should vanish because the
anti-fundamental
representation can be built by tensoring the antisymmetric
representation with itself ${1\over 2}(N-1)$ times. When $N$ is even
 string tension must exist. Note that \eqref{sigma} predicts
\bea
\label{as1}
\sigma = 2m\mu _{as} (N-2)(1-\cos {2\pi \over N-2})
\eea
which is not zero when $N$ is odd.

The resolution of the puzzle seems to be the following.
Non-Abelian charge can be static with respect to its spatial
location. However, its representation may change in time due to
emission or absorption of soft gluons (without cost of energy). 
Our semi-classical
description of the external charge as a c-number is insensitive to
this scenario. We need an extension of 
\eqref{external-charge} which takes into account the possibilities of
all various representations. One possible extension is  
\beq
\label{external-current}
j^a _{ext}=\delta ^{a3} k_{ext}(1+lN)(\delta(x^-+L)-\delta(x^--L))
\eeq
where $l$ is an arbitrary positive integer. This extension takes into
account the cases which correspond to
$1+lN$ charges multiplied in a symmetric way.   
The resulting string tension is 
\beq
\sigma =  m \mu _R \
\sum _i \left ( 1-\cos (4\pi \lambda _i  {k_{ext}  \over k_{dyn}}(1+lN)
  )\right ),
\eeq
which includes the arbitrary integer $l$. What is the value of $l$
that we should pick ?

The dynamical charges are attracted to the external charges in such a way
that the total energy of the configuration is minimal. Therefore the
value of $l$ which is needed, is the one that guarantees minimal string
tension.

Thus the extended expression for string tension is the following 
\beq
\sigma =  \min _l \left \{ m \mu _R \
\sum _i \left ( 1-\cos (4\pi \lambda _i  {k_{ext}  \over k_{dyn}}(1+lN)
  )\right ) \right \}\label{sigma2}
\eeq

In the case of dynamical antisymmetric charges and external
fundamentals and odd $N$, $l={1\over 2}(N-3)$ gives zero string
tension. When $N$ is even the string tension is given by \eqref{as1}.

The expression \eqref{sigma2} yields the right answer in some
other cases also, like the case of
dynamical charges in the symmetric representation. The bosonization
for this case can be derived in a similar way to that of the
antisymmetric representation, and $T^3$
is given by
\beq
T^3_{symm} =  diag (1,0,-1,
\underbrace{{1\over 2},-{1\over 2},{1\over 2},-{1\over 2},...,{1\over
    2},-{1\over 2}}_{(N-2)\,\,\, doublets},0,0,...,0), 
\eeq
 and therefore $k=N+2$. When the external charges transform in the fundamental
representation and $N$ is odd, eq.\eqref{sigma2}
predicts zero string tension (as it should). When $N$
is even the string tension is given by 
\[
\sigma = 2m\mu _{symm} \left ((1-\cos {4\pi \over N+2})+(N-2)(1-\cos {2\pi \over N+2})\right)
\]

We discussed only the cases of the fundamental, adjoint,
anti-symmetric and symmetric representations, since we used
bosonization techniques which are applicable to a limited class of
representations\cite{GNO}.

\section{Supersymmetric Yang-Mills}

The same technique can be used to prove screening in \sym.
In this case  the action is \cite{ferrara}  
\beq
 \label{sym} S = \int d^2 x \ tr \left ( -{1\over 4e^2} F^2_{\mu\nu} +  i\bar
 \lambda\Dslash\lambda
 +{1\over 2} (D_\mu \phi )^2 - 2ie \phi \bar \lambda \gamma _5 \lambda
\right ) ,
\eeq
where  $A_\mu $ is the
 gluon field, $\lambda $ the gluino (a Majorana fermion) and $\phi $
 a pseudo-scalar, are the components of the vector supermultiplet and
transform as the adjoint representation of $SU(N_c)$. Also $D_\mu =
\partial _\mu - i[A_\mu,.]$.

The action \eqref{sym} is invariant under SUSY
\ber
&&  \delta A_\mu = -i e\bar \epsilon \gamma _5 \gamma _\mu  \sqrt{2} \lambda \\
&&  \delta \phi = - \bar \epsilon \sqrt{2} \lambda \\
&&  \delta \lambda \ = {1\over {2 \sqrt{2}}e} \epsilon \epsilon ^{\mu \nu}
  F_{\mu \nu} +{i \over {\sqrt{2}}} \gamma ^{\mu} \epsilon D_{\mu} \phi
\eer

We now introduce an external current. The external
source breaks explicitly supersymmetry. However, this breaking does
not affect our derivation.
 We assume a semi-classical quark anti-quark
pair which points in some direction in the algebra. Without loss of
generality this direction can be chosen as the '3' direction
('isospin'). The additional part in the Lagrangian is $-tr\ j_\mu
^{ext} A^\mu$ where $j_0 ^{a\ ext} = [C(R_{ext})] \delta ^{a3}(\delta (x+L) - \delta (x-L))$
and $[C(R_{ext})]$ is  a c-number which depends on the representation
of the external source, in analogy with $k_{ext}$ of Chapter 7. The
interaction term can be eliminated by a left-handed
rotation of the gluino field in the '3' direction (we are using a
spherical basis, and so we can perform appropriate complex
transformation also for real fermions)

\bea
 \lambda  \rightarrow \tilde \lambda = e^{i\alpha(x) {1\over 2}(1-\gamma_5 ) T^3} \lambda \\
  {\bar \lambda}  \rightarrow {\tilde {\bar \lambda }}  =
 \bar \lambda  e^{-i\alpha(x) {1\over 2}(1+\gamma_5 ) T^3} 
\eea
 $T^3$ is in the 3 direction of the adjoint representation 
\ber
 &&
T^3 = diag (\mu _1,\mu _2, ...,\mu _{N_c^2 -1}) \\
 &&       = diag (1,0,-1,
\underbrace{{1\over 2},-{1\over 2},{1\over 2},-{1\over 2},...,{1\over
    2},-{1\over 2}}_{2(N_c-2)\,\,\, doublets},\underbrace{0,0,...,0}_{(N_c-2)^2})
\eer

 The chiral
rotation introduces an anomaly term $tr\ {\alpha(x) T^3 \over 4\pi} F$,
which is used to cancel the external charges. 

The choice $\alpha (x) = 2\pi {C(R_{ext}) \over N_c} (\theta (x+L) -\theta
   (x-L))$ leads to an action which is similar to the
   original \eqref{sym}, but has a chiral rotated term. The
   information of the external source is now transformed into a rotation angle.

The terms which are relevant to the computation of the string tension
are those which appear in the interaction Lagrangian. In this case, it
is the gluino pseudo-scalar term
\beq
 tr \ 2i \phi \bar \lambda \gamma _5 \lambda \rightarrow 
 tr \ 2i \phi \tilde {\bar \lambda}  \gamma _5 {\tilde \lambda}
\eeq

Let us see how this change influences the Hamiltonian vacuum energy. In the
original theory, without the external source, the Hamiltonian $H_0$
has no v.e.v., since the theory is supersymmetric and
$H_0 \sim Q^2$ (where $Q$ is the supercharge). In particular it means that there is no
$<tr\ \phi {\bar \lambda}  \gamma _5 { \lambda}>$
condensate\footnote{We are grateful to D.J. Gross for a discussion
  about this issue.}. The
reason is the following. The light-cone Hamiltonian density of the system is
given by 
\beq
{\cal H}= tr \ e^2 ({1\over \partial _-} j^+ )^2 + tr \ 2ie \phi \bar \lambda \gamma _5 \lambda 
\eeq
where $j^+$ denotes the total, scalar and gluino, current which
couples to the gauge field.  
SUSY implies that $<{\cal H}>=0$. In addition we may use the Feynman-Helman
theorem
\beq 
0=<{\partial {\cal H} \over  \partial e}> =   tr \ 2e ({1\over \partial _-} j^+ )^2 + tr \ 2i \phi \bar \lambda \gamma _5 \lambda 
\eeq
Thus, there are no non-trivial condensates
\bea
 &&  <tr \ e^2 ({1\over \partial _-} j^+ )^2>=<tr \ F^2> =0 \\
&&   <tr \ 2ie  \phi \bar \lambda \gamma _5 \lambda >= 0
\eea   

Note that we
assumed that SUSY is not broken dynamically. The numerical analysis
of\cite{ALP} indicates that this is indeed the case.

Let us compute the Hamiltonian density of the rotated theory. In the
regime $-L < x < L$
\beq
<{\cal H}> = 2ie <tr\ \phi \tilde {\bar \lambda}  \gamma _5 {\tilde
  \lambda}>
\eeq
By using the fact that $T^3$ is diagonal, and the vacuum state is
color symmetric, we get
\bea
\lefteqn{ <tr\ \phi \tilde {\bar \lambda}  \gamma _5 {\tilde
    \lambda}>=} \\
&&
 {1\over {N_c^2-1}} \sum _a \cos (\alpha \mu _a) <tr\ \phi {\bar
   \lambda}  \gamma _5 { \lambda}>
-i {1\over {N_c^2-1}} \sum _a \sin (\alpha \mu _a) <tr\ \phi {\bar
   \lambda}  { \lambda}> \nonumber ,
\eea
where $\alpha = \lim _ {L \rightarrow \infty} \alpha (x)$.
The first term on the right hand side vanishes since as argued before $<tr\ \phi {\bar
  \lambda}  \gamma _5 { \lambda}>=0$, and the second term vanishes
since the isospin eigenvalues, $\mu _a$, come in pairs of opposite
signs.

Thus $<{\cal H}>=0$ and the string tension is zero. 

Note that though we used the classical expression for the external current
and the effective Hamiltonian may include other terms, these terms
cannot change the value of the string tension. It is so because this
theory contains only one dimension-full parameter, the gauge coupling
$e$, and therefore the string tension is some number times $e^2$. We
showed that this number is zero and higher terms in $e$
which may appear in the effective Hamiltonian cannot affect the string tension.

The meaning of the last result is that a quark anti-quark pair
 located at $x=\pm \infty$ generate a linear potential. Physically, it is a consequence
of infinitely many adjoint fermions and scalars which are produced
from the vacuum, as there is no mass gap,  that are attracted to
the external source, form a soliton in the fundamental representation
 and result in screening it. A complementary argument\cite{FS,AS} is that due to
loop effects, the intermediate gauge boson acquires a mass $M^2 \sim e^2 N_c$, which
leads to a Yukawa potential between the external quark anti-quark pair. 

The above result can be generalized to theories with extended
supersymmetry and additional massive or massless matter content.

We argue that any supersymmetric gauge theory in two dimensions is
screening.
Technically, the reason is that the gluino is coupled to other fields
in such a way that $<{\cal H}>=0$ (guaranteed if SUSY is not
broken dynamically) and therefore there are no non-trivial chiral condensates. 
However, since the string tension is proportional to chiral
condensates, SUSY leads to zero string tension. 
Physically, it follows from the fact that the gluino is an adjoint
{\em massless} fermion. Since it does not acquire mass, external
sources are screened, as in the non-supersymmetric massless model.
Recently the spectrum of various supersymmetric models was
derived. The shape of the spectrum confirms our prediction\cite{ALP,A2}. 

In fact, the essential
requirement for a screening nature of the type argued above, is to
have among the charged particles at least one massless particle whose
masslessness is protected by an unbroken symmetry. The symmetry can be
gauge symmetry combined with supersymmetry or chiral symmetry.

\Acknowledgements
We thank  D.J. Gross and A. Zamolodchikov for illuminating
discussions.

The work of J.S. is supported in part by the Israel Science
Foundation, the US-Israel Binational
Science Foundation and the Einstein Center for Theoretical Physics at
the Weizmann Institute.
The work of A.A. is supported in part by the Einstein Center for Theoretical Physics at
the Weizmann Institute.

\appendix
\section{Appendix - The external field}
We give here a detailed derivation of the external quark anti-quark field
\eqref{external-charge} for $N>2$ and with $2\pi$ in exponent for $N=2$.

For the case of external charges in a real representation the $u$
field can chosen to point in some special direction in the $SU(N)$ algebra which we take to
be '3', namely $u=\exp -i T^3 \phi$. For external
charges in a complex representation one has to dress this ansatz with
a baryon number part, namely  $u=\exp -i\chi \exp
-iT^3\phi $. Let us view the external source as the limit of a
dynamical variable with very large mass. Let us choose the gauge
$A_-=0$. Then we can take $A_+=a_+ T^3$, as the other directions do not
couple. The Lagrangian for the real case takes
the form
\bea
&& {\cal L} = {k\over 8\pi} (\partial _- \phi)( \partial _+ \phi ) +
{1\over 2e^2} (\partial _- a_+)^2 +\\
&&  +M^2
\sum _i \cos \lambda _  i \phi + {k\over 4\pi} \partial _-\phi a_+
\nonumber ,
\eea
where $k$ is the level 
 and $\lambda _i$ the isospin entries of the diagonal sub $SU(2)$ generator $T^3$.

The equations of motion for the matter and gauge fields are
\bea
&& \label{phi} {k\over 4\pi} \partial _ - \partial _+ \phi + M^2 \sum _i \lambda _i
\sin \lambda _i \phi + {k\over 4\pi} \partial _- a_+=0 \\
&& \label{A} \partial _- ^2 a_+ = e^2 {k\over 4 \pi}  \partial _- \phi
\eea
Integrating \eqref{A} with zero boundary conditions and substituting in \eqref{phi} we obtain 
\beq
 {k\over 4\pi} \partial _ - \partial _+ \phi + M^2 \sum _i \lambda _i
\sin \lambda _i \phi + e^2 {({k\over 4\pi})}^2 \phi =0
\eeq
Let us assume a solution for $\phi$ which describes an infinitely
heavy light-cone static quark anti-quark system
\beq
\phi = \alpha \left (\theta (x^- +L)-\theta(x^- -L)\right ),
\eeq
where $\alpha$ is a yet unknown coefficient.

For the region $-L<x^-<L$ we obtain
\beq
\label{alpha}
M^2 \sum _i \lambda _i \sin \lambda _i \alpha +  e^2 {({k\over
    4\pi})}^2 \alpha =0
\eeq
When $M^2 \gg e^2$ the solution for $\alpha$ is of the form 
\beq 
\alpha = 4\pi n + \epsilon,
\eeq
where $n$ is integer (we will pick the minimal $n=1$ possibility) and
the small parameter $\epsilon$ is determined by the substitution in \eqref{alpha}
\beq
M^2 \sum _i \lambda _i ^2 \epsilon  +  e^2 {({k\over
    4\pi})}^2 4\pi \approx 0
\eeq  
 Thus $\alpha$ is given by
\beq 
\alpha = 4\pi - {e^2 \over M^2}  {({k\over
    4\pi})}^2 {4\pi \over \sum _i \lambda _i ^2} +
O\left ({({e^2\over M^2})}^2 \right)
\eeq
In the limit $M^2 \rightarrow \infty$, $u$ is 
\beq
u=\exp -i4\pi \left ( \theta (x^-+L )-\theta
  (x^--L )\right ) T^3
\eeq 
When $u$ is in a complex representation, namely $u$ is represented by 
$u=\exp -i\chi \exp -iT^3\phi$, we find by repeating the above
derivation the following expression
\bea
\lefteqn{u=
\exp -i2\pi \left ( \theta (x^-+L )-\theta
  (x^--L )\right ) \times} \\ &&
 \times \exp -i4\pi \left ( \theta (x^-+L )-\theta
  (x^--L )\right ) T^3 \nonumber, 
\eea
for $U(N>2)$ and
\bea
\lefteqn{u=
\exp -i\pi \left ( \theta (x^-+L )-\theta
  (x^--L )\right ) \times} \\ &&
\times \exp -i2\pi \left ( \theta (x^-+L )-\theta
  (x^--L )\right ) T^3 \nonumber, 
\eea
for $U(2)$.
Note that the $SU(2)$ part has a $2\pi$ prefactor. The reason is that
for $SU(2)$, it is the only case where the adjoint does not contain
isospin ${1\over 2}$.

\end{document}